\documentclass[11pt,twoside]{article}

\def\approxgt{\mathrel{\hbox{\rlap{\lower.55ex \hbox {$\sim$}}
        \kern-.3em \raise.4ex \hbox{$>$}}}}
\def\approxlt{\mathrel{\hbox{\rlap{\lower.55ex \hbox {$\sim$}}
        \kern-.3em \raise.4ex \hbox{$<$}}}}
\usepackage{asp2006}
\usepackage{epsf}
\usepackage{psfig}
\usepackage{lscape}

\begin{document}
\title{``Warm absorbers'' and ``-mirrors'': one and the same gas?}   
\author{Matteo Guainazzi$^1$, Stefano Bianchi$^{2,1}$}   
\affil{$^1$European Space Astronomy Center of ESA, Apartado 50727, E-28080 Madrid, Spain \\ $^2$Dipartimento di Fisica ``E.~Amaldi'', Universit\`a Roma Tre, Via della Vasca Navale 84, I-00146 Roma, Italy}    

\begin{abstract} 
We review the main properties of ``warm mirrors'' in obscured AGN, and
discuss whether the scattering gas could be also responsible
for ``warm absorbers'' commonly observed in unobscured AGN.
\end{abstract}



``X-ray obscured'' (generally type~2)
AGN are not totally X-ray silent at energies below their soft X-ray
photoelectric cut-off.
The advent of broad-band moderate resolution X-ray spectroscopy led to
the discovery of soft X-ray emission, in excess above the extrapolation
of the obscured nuclear continuum \cite{turner97a}.
This spectral feature
is present almost ubiquitously in samples of nearby Seyfert~2
galaxies \cite{guainazzi05}.
The dramatic improvement
in the spatial and spectral resolution provided by the scientific payload
on-board {\it Chandra} and XMM-Newton has allowed us to
understand, almost unambiguously, the nature of this spectral component.
Obscured AGN soft X-ray spectra are dominated by
emission lines from He- and H-like transitions of elements from Carbon
to Neon, as well as by L-shell transitions from Fe{\sc xvii} to
Fe{\sc xxi} \cite{guainazzi06}. ``Narrow'' (intrinsic
width, $\sigma \approxlt 10$~eV)
Radiative Recombination Continuum features
from C{\sc v}, O{\sc vii} and O{\sc viii} are unambiguous signatures of
photoionization. The
comparatively large intensity of higher order transitions - when compared
to their K$_{\alpha}$ - indicates that resonant scattering also plays
an important role in the ionization balance \cite{kinkhabwala02}.
This prevents standard
spectral diagnostics \cite{gabriel69,porquet00} from being usable. Nonetheless,
spectra of obscured AGN and of starburst galaxies are systematically
different, once diagnostic parameters unaffected by resonant scattering are
considered \cite{guainazzi06}. This does not impede that in a few type~2
sources \cite{levenson05} - often low-luminosity AGN
\cite{jimenezbailon03} -
the X-ray Spectral Energy Distribution is dominated by stellar processes.

Furthermore, the morphology of the soft X-ray emitting gas exhibit a
striking coincidence with high-resolution O[III] images on scales
as large as $\sim$0.1--1~kpc \cite{young01,bianchi06}.
Simple mono-phase photoionization models reproduce well the observed
X-ray to optical luminosity ratio \cite{bianchi06}.
Successful models yield an almost
constant ionization parameter across the whole extension of the gas, hence
a radial dependence of the electronic density $n_e \propto r^{-\beta}$,
with $\beta \simeq$1.8--2.0. The latter
results are in good agreement with those derived
from spatial-resolved optical spectroscopy of the Narrow Line Regions
(NLRs) \cite{kraemer00,bradley04,collins05}.
The possibility that the soft X-ray extended
emission is mainly
powered by ``local'' photoionization, due to gas heated by mechanical
shocks in the interaction between a
jet and the interstellar medium, cannot in principle be ruled out.
However, 
the discovery of this close connection between soft X-ray emitting gas
and NLRs again points to AGN-photoionization as the main
physical mechanism responsible for the ionization balance.

The physical properties of these X-ray {\bf 
``warm mirrors''} are relatively poorly
constrained. Detail photoionization models applied to the best quality
spectra indicate:

\begin{itemize}

\item column densities in the range 10$^{21-22}$~cm$^{-2}$
\cite{sako00,sambruna01,bianchi02,kinkhabwala02}. 

\item a wide range of ionization parameters ($\xi = L/n_e r^2$,
where $L$ is the ionizing luminosity):
$\log (\xi) = 0$--3. The soft X-ray spectra
we measure are probably due to a mixture of contributions from different gas
phases \cite{sambruna01,bianchi02}\footnote{It
should be born in mind that X-ray spectroscopic
slits typically encompasses a few kilo-parsecs
around the nucleus even in the closest AGN}.
The detection of fluorescent emission lines from highly ionized {\it
iron} \cite{bianchi05} confirms this hypothesis

\item turbulent
velocities of the order of a few hundreds km~s$^{-1}$, in the
very few cases where these measurements are possible
\cite{kinkhabwala02,guainazzi06}

\end{itemize}

The existence of an electron-dominated
scattering plasma filling the torus axial region is one of the
basic ingredient of standard Seyfert unification scenarios, invoked
to explain wavelength-independent polarization
of optical broad lines in polarimetric measurements of some type~2
AGN \cite{antonucci85,tran95}.

In a large fraction - probably
as high as 50\%  - of ``X-ray
unobscured'' (generally type~1)
AGN the high-energy nuclear radiation is transmitted through
photoionized gas, the so called {\bf ``warm absorber''}.
Since their discovery \cite{halpern84} and early characterization
\cite{reynolds97,george98}, warm absorbers represent a fundamental
ingredient of any AGN structure model.
An Occam's razor argument may suggest that ``warm absorbers'' and
``warm mirrors'' are actually one and the same gaseous system.
Again, 
our understanding of the physical and dynamical properties of the
absorbing gas
has dramatically improved with the advent of X-ray high-resolution
spectrometers. After a recent study by Blustin et al. (2005) on a sample of
23 nearby AGN, one can summarize the properties of the warm absorbing gas
as follows: {\it i)}
average ionization parameter covering a wide range
($\log (\xi) = 0$--3)\footnote{Actually, the range of
ionization parameters is even wider than that, if one include
in-/outflows traced by
absorption from He- and H-like iron \cite{pounds03,dadina05},
as well as spectral signatures of low-ionization gas such as the
Unresolved Transition Arrays of M-shell iron \cite{behar01}};
{\it ii)} total column densities in the range 10$^{21-22}$~cm$^{-2}$;
{\it iii)} typical outflowing velocities of a few hundreds km~s$^{-1}$,
except for ``iron'' outflows, whose velocities can be up to one order of
magnitude larger. The basic gas physical parameters are therefore
largely coincident between ``warm absorbers'' and ``warm mirrors''.

\begin{figure}
\begin{center}
\psfig{file=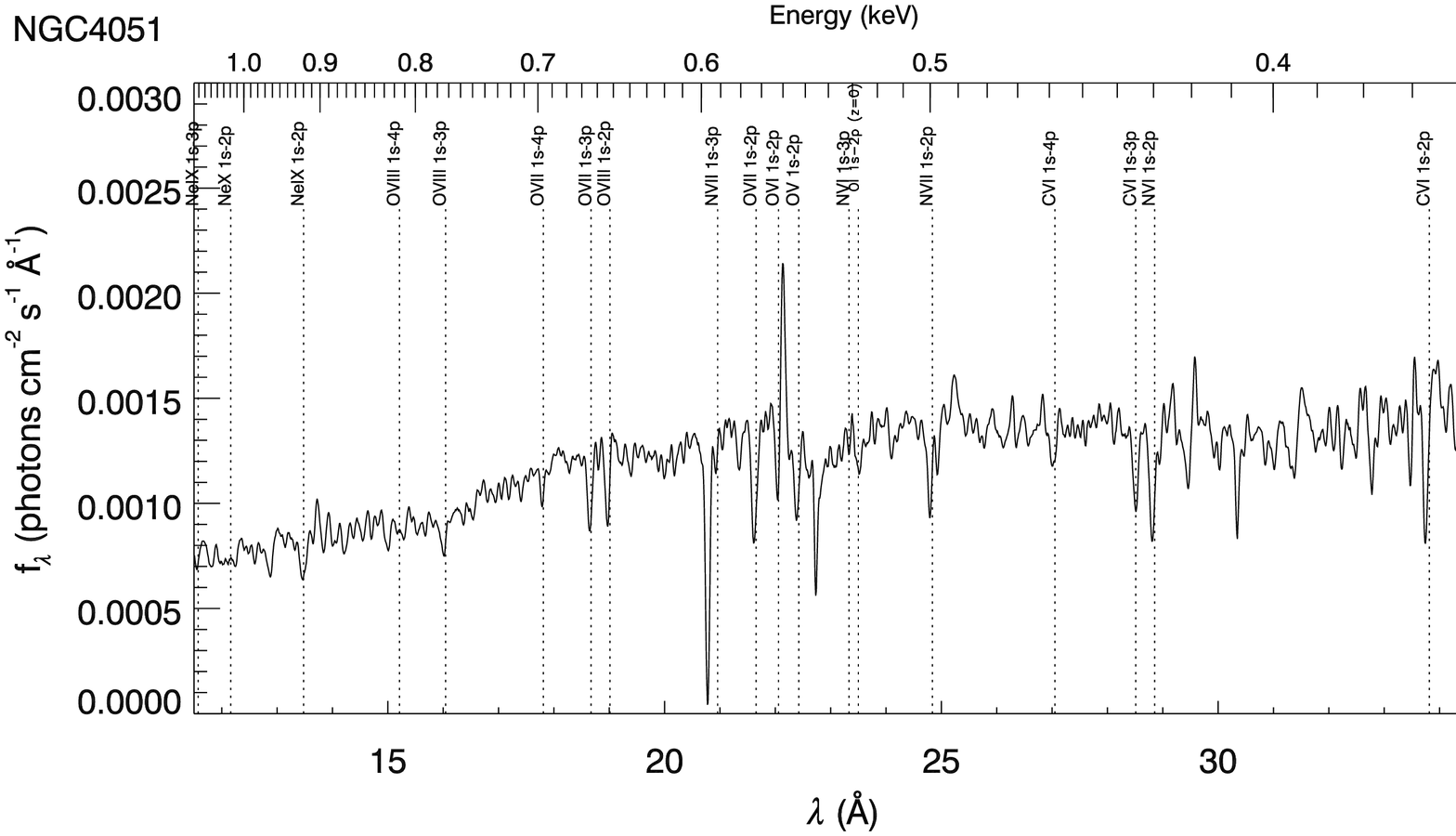,height=60mm,width=120mm,angle=90}
\vspace{1.0cm}
\psfig{file=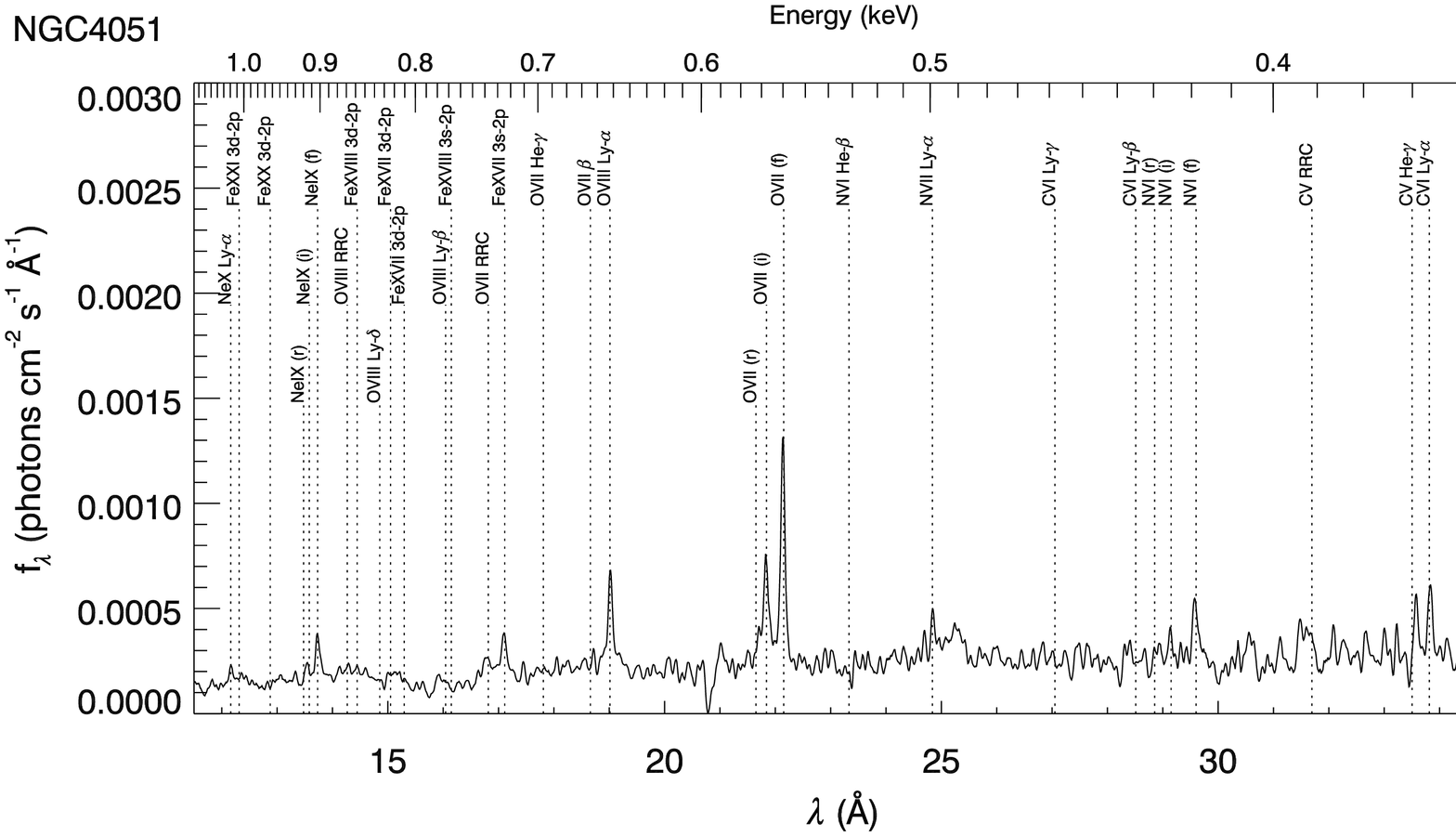,height=60mm,width=120mm,angle=90}
\end{center}
\caption{XMM-Newton RGS spectra
of the Narrow Line Seyfert~1 Galaxy NGC~4051 during a normal flux state
(May 2001; {\it top panel}) and a low flux state ({\it bottom panel}).
The former exhibits typical features of a ``warm absorber'';
the latter, taken 18 months later, is almost indistinguishable
from the ``warm mirror''-dominated soft X-ray spectrum of
an obscured AGN. The deep feature at $\lambda \simeq 20.80$\AA\
is instrumental. The main transitions observed typically in
warm absorbers and warm mirrors are labeled for reference.
See also Pounds et al. 2004.}
\label{fig1}
\end{figure}

The exact geometry of the warm absorber is unknown, although dynamical
constraints are consistent with it
originating as a radiation-driven
high-velocity outflow in accretion disk instabilities
(Krongold et al. 2005; Nicastro, this volume), and
propagating up to typical
NLR distances \cite{kraemer06}.
These uncertainties prevent
firm statements on the identity between warm absorbers and warm mirrors
from being drawn.
Interestingly enough, however, there is at least one known
source, exhibiting a transition from an absorber- into a 
mirror-dominated soft X-ray spectrum.
NGC~4051 is one of the most dramatically X-ray variable AGN in the local
Universe. It is a $L_X \sim 10^{42}$~erg~s$^{-1}$ Narrow Line Seyfert~1
Galaxy, which occasionally exhibits extreme
low flux states. During one of these states, XMM-Newton
observed a spectrum practically indistinguishable from an highly obscured
AGN \cite{pounds04}, at odds with the standard ``warm absorbed'' spectrum
observed during
normal flux states (see Fig.~\ref{fig1}).


\end{document}